\documentstyle[epsf,prb,floats,twocolumn,aps]{revtex}

\begin{document}

\baselineskip 11.5 pt

\title{Modelling the Nonlinear High-Frequency Response of a Short
Josephson Junction under Two-Frequency Irradiation}

\author{Anton V. Velichko$^{1,2}$ and  Adrian Porch$^1$
\\ {\small$^1$School  of Electronic \& Electrical Engineering, University
of Birmingham, UK}
\\ {\small $^2$Institute of Radiophysics \& Electronics of NAS, Ukraine}}

\maketitle

\abstract{%
The nonlinear response of a short Josephson Junction (JJ), being irradiated
simultaneously with two high-frequency signals, has been studied in the
framework of the nonlinear Resistively-Shunted Junction (RSJ) Model. One of
the signals, hereafter referred to as ``probe signal", has a small
amplitude $I_{pr}<I_c$ ($I_c$ is the critical current of the JJ) and
frequency $f_{pr}$, and is used to monitor the response of the
junction to the other high-power signal with amplitude $I_{pm}$ and
frequency $f_{pm}$, hereafter referred to as ``pump signal". Varying the
frequency ratio $f_{pm}/f_{pr}$ from 0.5 to 100, and the current
amplitude of the probe signal from 0.01 to 0.9 of $I_c$, we found that
the dependence of the junction impedance at the frequency $f_{pr}$,
$Z_s^{f_{pr}}$, versus $I_{pm}$ preserves its general features,
independent of $f_{pm}/f_{pr}$ and $I_{pr}/I_c$ values. At the
same time, some particular features, like negative values of
$Re(Z_s^{f_{pr}})$ and ``fine" structure of the steps in
$Z_s^{f_{pr}}(I_{pm})$ are observed for $f_{pm}/f_{pr}<1$
and for particular values of $I_{pr}/I_c$. In general, the behavior of
$Z_s^{f_{pr}}(I_{pm})$ is rather different from that predicted by the
nonlinear RSJ model for a short JJ in the regime of single-frequency
irradiation, when one and the same signal plays the roles of the pump and
the probe signals simultaneously. Possible applications of the model are
briefly discussed.}

\section{Introduction}

The high-frequency nonlinear response of a Josephson Junction (JJ) is of
significant interest because JJs are necessary units of almost all active
microwave and millimeter (mm) wave devices~\cite{Likhbook}, and because
weak links, which are likely to be present even in the highest quality
samples of high-$T_c$ superconductors (HTS), can be modelled as
JJs~\cite{VelCher}. The RSJ model is often used to simulate the
characteristics of JJ-based devices~\cite{Likhbook}, and was shown to give
a good agreement with experimental data on point-contacts and microbridges
in the microwave and mm wave ranges, where the capacitance of the junction
can be neglected.

\begin{figure}[t]
\def\epsfsize#1#2{0.47#1}
\centerline{\epsfbox{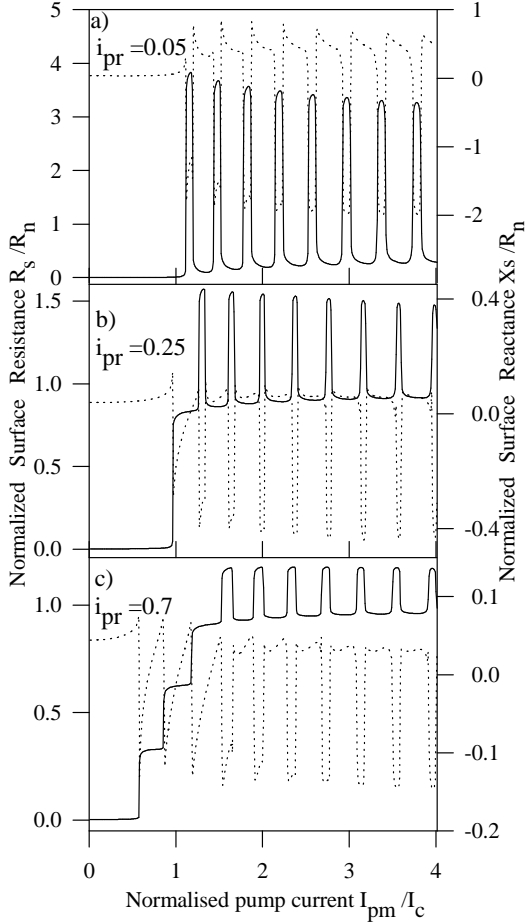}}
\caption{Normalised surface resistance $R_s/R_n$ (solid line)
and surface reactance $X_s/R_n$ (broken line)
for a short Josephson junction, as a function of normalised
pump current amplitude $i_{pm}=I_{pm}/I_c$, simulated within the
framework of the RSJ model for the two-frequency case. Here, the frequency
ratio $\Omega_{pm}/\Omega_{pr}=2$, and the probe current $i_{pr}$ is that
given in each of the figures.}
\label{fig1}
\end{figure}

\begin{figure}[t]
\def\epsfsize#1#2{0.47#1}
\centerline{\epsfbox{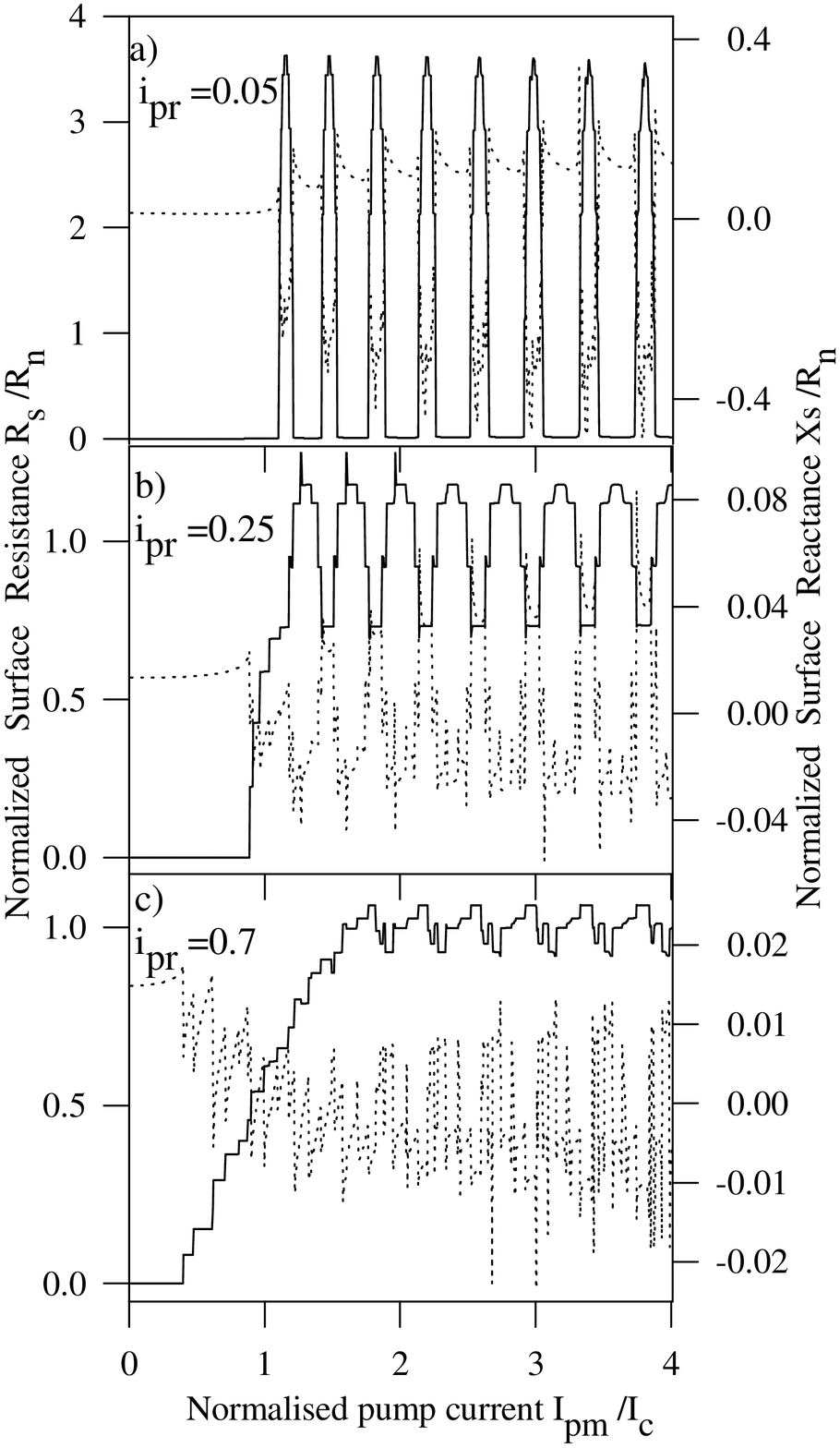}}
\caption{$R_s/R_n$ and $X_s/R_n$ for a short Josephson junction,  as a
function of $i_{pm}=I_{pm}/I_c$, simulated within the framework of the RSJ
model for the two-frequency case. Here, the frequency ratio
$\Omega_{pm}/\Omega_{pr}=10$, and the probe current $i_{pr}$ is that given
in each of the figures.}
\label{fig2}
\end{figure}

\begin{figure}[t]
\def\epsfsize#1#2{0.47#1}
\centerline{\epsfbox{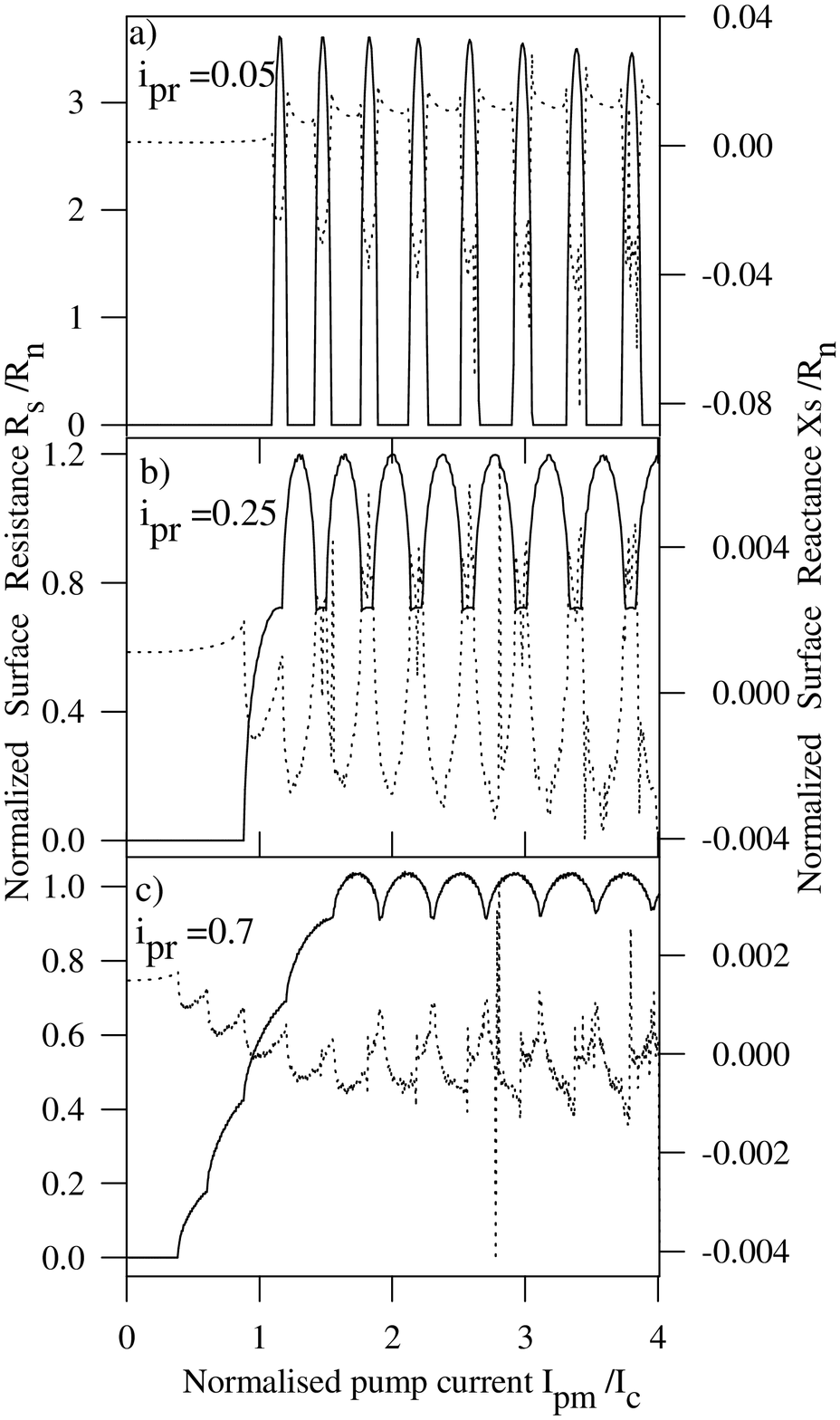}}
\caption{$R_s/R_n$ and $X_s/R_n$ for a short Josephson junction,  as a
function of $i_{pm}=I_{pm}/I_c$, simulated within the framework of the RSJ
model for the two-frequency case. Here, the frequency ratio
$\Omega_{pm}/\Omega_{pr}=100$, and the probe current $i_{pr}$ is that
given in each of the figures.}
\label{fig3}
\end{figure}

\begin{figure}[t]
\def\epsfsize#1#2{0.47#1}
\centerline{\epsfbox{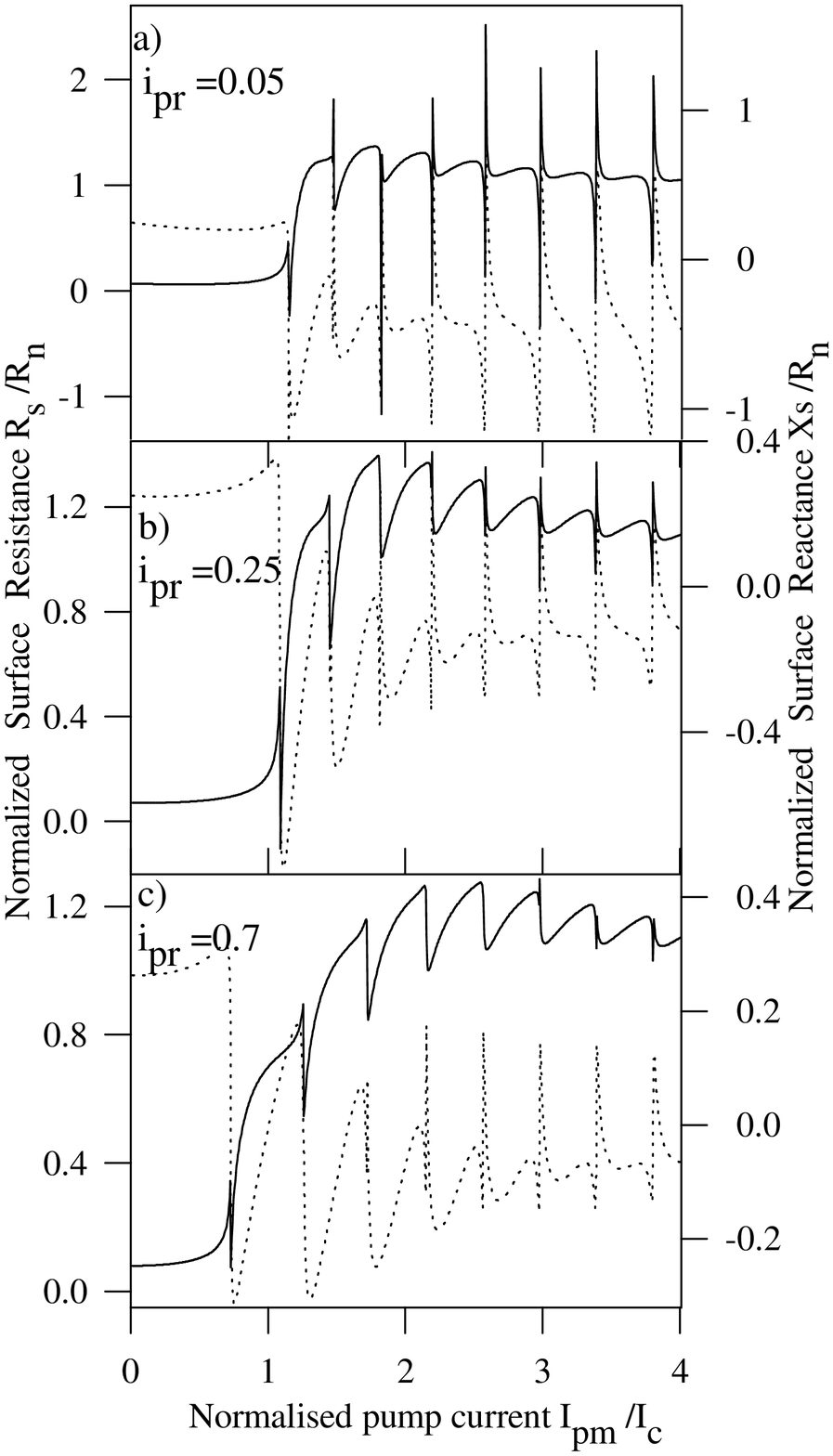}}
\caption{$R_s/R_n$ and $X_s/R_n$ for a short Josephson junction,  as a
function of $i_{pm}=I_{pm}/I_c$, simulated within the framework of the RSJ
model for the two-frequency case. Here, the frequency ratio
$\Omega_{pm}/\Omega_{pr}=0.5$, and the probe current $i_{pr}$ is that
given in each of the figures.}
\label{fig4}
\end{figure}

In the present paper we report simulation of the surface impedance of
a JJ at the frequency of the low amplitude signal, hereafter referred to as
``high-frequency response", as a function of the current amplitude of the
other elevated-power high-frequency signal. This case can be considered as
a model of a microwave-biased electromagnetic radiation detector, which
has been shown to have an improved sensitivity and noise figures when
compared with dc-biased detectors~\cite{Petrov,Eldum}. The other possible
implication of the model is modeling the performance of microwave
parametric amplifiers and Josephson mixers~\cite{Likhbook}. In addition,
the above model can also describe the nonlinear microwave response of
superconducting weak links, which is often investigated using the
so-called ``pump-probe" technique~\cite{VelCher,Velich1,Leviev}. This
technique is of a particular interest because it allows one to modulate or
to pulse the powerful microwave signal, whilst measuring the surface
impedance of the sample with the help of the other low-amplitude
continuous wave microwave signal at a different frequency. In such a way,
the pump-probe method avoids substantial heating effects and allows the
study of intrinsic nonlinear phenomena in superconductors. However, using
the pump-probe technique, one has to know how the nonlinear surface
impedance, measured at the pump frequency, relates to that measured at the
frequency of the probe signal, with respect to which the superconductor is
in the linear regime.  Although the model proposed in this paper cannot
be considered as a model of the nonlinear response of HTS, an extension of
the model to the case of a 2D JJ array with a random distribution of
$I_c\cdot R_n$-products (as recently proposed by Herd et al.\ for the
single-frequency case~\cite{Herd}) would allow the description of
experiments on HTS thin films in the pump-probe
regime~\cite{VelCher,Velich1,Leviev}.

\section{Numerical Simulation}

In the case of two signals applied to a short JJ,
the sine-Gordon Equation (which describes time dependence of the order
parameter phase difference $\varphi$ across the junction) in dimensionless
form can be written as follows~\cite{VanDuz}

\begin{equation}
\frac{d\varphi}{d\tau} = i_{pm}\sin(\Omega_{pm}\tau) +
i_{pr}\sin(\Omega_{pr}\tau) - \sin\varphi,
\label{equ.1}
\end{equation}
\noindent%
where $\tau=\beta t$,  $i_{pm} = I_{pm}/I_c$, $i_{pr} = I_{pr}/I_c$,
$I_{pm}$ and $I_{pr}$ current amplitudes of the pump and probe signals
respectively, $\omega_{pm}=2\pi f_{pm}$ and $\omega_{pr}=2\pi f_{pr}$ the
corresponding circular frequencies, $I_c$ the critical current of the
junction, $\beta=2eR_nI_c/\hbar$, $\Omega_{pm}=\omega_{pm}/\beta$,
$\Omega_{pr}=\omega_{pr}/\beta$, and $R_n$ surface resistance of the
junction in the normal state. Equation (\ref{equ.1}) does not take into
account the effect of the junction capacitance, which can be neglected at
microwave frequencies.

Generally, the solution of (\ref{equ.1}) is not necessarily periodic
with time. Only in the case when the ratio
$\Omega_{pm}/\Omega_{pr}$ is an integer is the solution of (\ref{equ.1})
periodical with a period equal to the least common multiple (LCM) of the
pump and the probe signal periods. If we expand the phase derivative
$\dot{\varphi}$ into a Fourier series with respect to time over the LMC
period of the two signals, we obtain coefficients which couple the
voltage $\sim\dot{\varphi}$ to the current $I_{r\!f}$. If then we single
out the series' terms at the probe frequency, we obtain the surface
impedance at the relevant frequency as follows

\begin{equation}
Z_s = R_s+jX_s = \lim_{n \to \infty}  \frac{\Omega R_n}{\pi i_{pr}n}%
\int\limits_0^{2\pi n/\Omega} \dot{\varphi}(\tau)\exp(j\Omega_{pr}\tau)
d\tau,
\label{equ.2}
\end{equation}
where $\Omega=2\pi/T$, and $T$ is the LCM of the pump and probe signal
periods. Because one can assume different initial conditions for
(\ref{equ.1}), its solution is not strictly periodic with $T$, and hence
in (\ref{equ.2}) integration over a few periods is required to get an
appropriate convergency of the results.

The parameters used for simulation are as follows:
$R_n=10^{-3}~\Omega$, $I_c=0.5$~A, $f_{pm}=3.6\cdot 10^{10}$~Hz,
$f_{pr}=$(0.036--7.2)$\cdot 10^{10}$~Hz,
$\Omega_{pm}=2\pi\hbar f_{pm}/(2e I_c R_n) =0.141$.

Results of the simulation for different ratios of
$\Omega_{pm}/\Omega_{pr}$ varied from 0.5 to 100, and for different values
of the probe current amplitude $I_{pr}$ (0.05, 0.25, and 0.7) are plotted
in Fig.~\ref{fig1}, Fig.~\ref{fig2}, Fig.~\ref{fig3} and Fig.~\ref{fig4}.
For $\Omega_{pm}/\Omega_{pr}>1$ (Fig.~\ref{fig1}a, Fig.~\ref{fig2}a and
Fig.~\ref{fig3}a) at low $i_{pr}$ (0.05), the surface resistance
$R_s^{f_{pr}}$ as a function of $I_{pm}$ is a sequence of peaks of about
the same height with a ``pedestal'' level approximately equal to the value
of $R_s^{f_{pr}}$ in the linear regime. This picture is rather different
from that expected for the single-frequency situation. In the latter case,
the surface resistance $R_s$ has a staircase-like form, starting to
increase rapidly for $i_{pm}>1$, and gradually approaching the $R_n$ value
as $i_{pm}\rightarrow \infty$. With regards to the surface reactance $X_s$,
for the single-frequency case it oscillates around zero with
amplitude decaying with increased $i_{pm}$ (see, e.g.\ ,~\cite{Herd}).
However, in the two-frequency regime, no obvious decay of the peaks
amplitude in $X_s$ up to $i_{pm}=4$ is observed. In addition, every
oscillation peak seen in $X_s(i_{pm})$ in the single-frequency regime
translates into a peak with a complicated structure, containing many
upward and downward minor peaks of smaller amplitude. With increased
$i_{pr}$ the major peaks in  $X_s^{f_{pr}}(i_{pm})$ broaden, and more
complicated ``fine'' structure of minor peaks develops (see
(Fig.~\ref{fig1}b,c, Fig.~\ref{fig2}b,c and Fig.~\ref{fig3}b,c)). As far as
$R_s^{f_{pr}}$ is concerned, an increase of $i_{pr}$ leads to the
appearance of steps in $R_s^{f_{pr}}(i_{pm})$, similar to those
seen in $R_s(i_{pm})$ for the single-frequency regime. The higher
$i_{pr}$, the more steps are observed in $R_s^{f_{pr}}(i_{pm})$, before it
levels off and starts to oscillate near some average value around unity
for $i_{pm}\gg 1$.

Contrary to the case of $\Omega_{pm}/\Omega_{pr}>1$, when the major peaks
in $R_s^{f_{pr}}(i_{pm})$ are almost symmetrical with respect to a
vertical line drawn through the middle of their width, in the case of
$\Omega_{pm}/\Omega_{pr}<1$ these peaks are obviously asymmetric  (see
Fig.~\ref{fig4}). Another distinctive feature for this case are
discontinuous double-peak structures (upward peak followed by downward
peak) at the beginning and the end of every major peak. One further
significant difference of $R_s^{f_{pr}}(i_{pm})$ in the case of
$\Omega_{pm}/\Omega_{pr}<1$, as compared to the case with
$\Omega_{pm}/\Omega_{pr}>1$, is the appearance of regions with negative
values of $R_s^{f_{pr}}$. This means that under these particular
conditions the JJ can contribute energy to the external circuit, i.\ e.\
it works as a generator. This effect was theoretically predicted and
experimentally observed in JJs made of low-temperature superconductors,
and was called  ``the effect of nondegenerated single-frequency parametric
regeneration''~\cite{Likh41&22}. As the theoretical analysis showed,
this phenomena can be realised in any parametric element, a
reactive parameter of which can take negative values with changing
time~\cite{Likh18&30}.

All other features of $R_s^{f_{pr}}(i_{pm})$ for the case of
$\Omega_{pm}/\Omega_{pr}<1$, such as the appearance of steps, an increase
in their number, and a shift of the oscillatory part of the dependence to
higher $i_{pm}$ with increased $i_{pr}$, are similar to those seen in the
case of $\Omega_{pm}/\Omega_{pr}>1$. As far as $X_s^{f_{pr}}(i_{pm})$ is
concerned, features like asymmetry of the major peaks and discontinuous
double-peak structures are observed, similar to those present in
$R_s^{f_{pr}}(i_{pm})$.

\subsection{Implications for applications}
One of the possible applications of the two-frequency regime, simulated
in this paper, is a microwave-biased JJ detector. An advantage of this
regime is that the amplitude of the oscillation peaks (or, equivalently,
the impedance steps) can be made several times (up to a factor of
3--4) higher than the resistance of the junction in the normal-state,
especially at a small probe current (see Fig.~\ref{fig1}a,
Fig.~\ref{fig2}a and Fig.~\ref{fig3}a).  This should lead to enhanced
sensitivity of the detector, as compared to the single-frequency regime.
Despite that similar results (increased step heights) have been
also obtained for a dc-biased JJ~\cite{VanDuz}, a microwave-biased
detector was shown to benefit from reduced noise temperature and enhanced
responsivity as compared to the dc-biased one~\cite{Petrov,Eldum}.
Moreover, since the step amplitudes in the saturation regime ($I_{pm}\gg
1$) are almost independent of the pump current, such a detector will
possess amplitude-independent sensitivity.

In the case when $\Omega_{pm}/\Omega_{pr}<1$, the JJ can be used as a
parametric amplifier operating in the single-frequency nondegenerate
regime which, when compared to the self-pumped regime, was shown to give
a reduced noise temperature and a narrower frequency response of the
junction at the probe frequency~\cite{Likhbook}, allowing one to use high
quality factor resonators for matching the junction with the external
circuits.

\section{Conclusion}
The numerical simulation performed by us has shown that the nonlinear
high-frequency response $Z_s^{f_{pr}}$ of a short JJ in the regime of
two-frequency irradiation can be rather different from the surface
impedance $Z_s$ measured in the single-frequency regime. Depending on the
ratio of the pump to the probe frequencies, a number of new features in
$Z_s^{f_{pr}}(i_{pm})$ are predicted. Among them are the absence of the
steps in $R_s^{f_{pr}}(i_{pm})$ at low probe currents ($i_{pr}<0.05$);
persistent oscillations of $R_s^{f_{pr}}(i_{pm})$ around some average
value which tends to unity with increased $i_{pm}$; a multiple-peak
structure of $X_s^{f_{pr}}(i_{pm})$, which becomes more complicated with
increased ratio $\Omega_{pm}/\Omega_{pr}$; appearance of regions with
negative values of surface resistance in $R_s^{f_{pr}}(i_{pm})$ for the
case of $\Omega_{pm}/\Omega_{pr}<1$.

At the same time, there are some features which are similar to the those
in the single-frequency regime. These are the appearance of steps in
$R_s^{f_{pr}}(i_{pm})$ with increased $i_{pr}$, and the oscillation
of $X_s^{f_{pr}}(i_{pm})$ around a zero-$X_s^{f_{pr}}$ value.

The model presented here was shown to give a useful basis knowledge for an
application of the JJ as a microwave-biased detector of electromagnetic
radiation.

\section*{Acknowledgment}

A.V.V. thanks I.V.Yurkevich and A.S.Stepanenko for useful discussion.

\end{document}